\documentclass[]{raa}            % referee version: for submission
\usepackage{graphicx,times}
\usepackage{natbib}
\usepackage{subfigure}

\providecommand{\bm}[1]{\mbox{\boldmath$#1$\unboldmath}}

\begin{document}

   \title{Temporal  Variation  of the  Hemispheric Solar Rotation
 % $^*$
%\footnotetext{\small $*$ Supported by the National Natural Science Foundation of China.}
}

 \volnopage{ {\bf 20xx} Vol.\ {\bf 9} No. {\bf XX}, 000--000}
   \setcounter{page}{1}

   \author{J.-L. Xie
      \inst{1,2}
   \and X.-J. Shi
      \inst{1,2}
   \and J.-C. Xu
      \inst{1,2}
   }
%% Here is an example of three authors come from different institutes.
%% For single author or all the authors from an institute, use "\inst{}" only

   \institute{National Astronomical Observatories/Yunan Observatory, Chinese Academy of Sciences,
             Beijing 100012, China; {\it xiejinglan@ynao.ac.cn}\\
%% Please give the E-mail address of the author, to whom future correspondence and
%% offprint requests will be sent.
        \and
             Graduate School of Chinese Academy of Sciences, Beijing~100863, China\\
\vs \no
   {\small Received [year] [month] [day]; accepted [year] [month] [day] }
}

\abstract{The daily sunspot numbers of the whole disk as well as the
northern and southern hemispheres from January 1, 1945  to December
31, 2010 are used to investigate the temporal variation of the
rotational cycle length
 through the  continuous  wavelet transformation analysis
method. The auto-correlation function analysis of daily hemispheric
sunspot numbers shows that the southern hemisphere rotates faster
than the northern hemisphere. The results obtained from the wavelet
transformation analysis are: there exists no direct relationship
between the variation trend of the rotational cycle length  and the
variation trend of solar activity in the two hemispheres; the
rotational cycle length of both hemispheres has no significant
period appearing at the 11 years, but has significant period of
about 7.6 years. Analysis concerning the solar cycle dependence of
the rotational cycle length shows that in the whole disk and the
northern hemisphere acceleration seems to appear before the minimum
time of solar activity. Furthermore, the cross-correlation study
indicates that the rotational cycle length of the two hemispheres
has different phases, and the rotational cycle length of the whole
disk as well as the northern and southern hemispheres also has phase
shifts with the corresponding solar activity. What's more, the
temporal variation of North-South (N-S) asymmetry of the rotational
cycle length is also studied; it displays the same variation trend
as the N-S asymmetry of solar activity  in a solar cycle as well as
in the considered time interval, and it has two significant periods
of 7.7 and 17.5 years. Moreover, the N-S asymmetry of the rotational
cycle length and the N-S asymmetry of solar activity are highly
correlated. It's inferred that the northern hemisphere should rotate
faster at the beginning of solar cycle 24.
  }

   \authorrunning{J.-L. Xie,  X. J. Shi \& J. C. Xu}            %author_head in even pages
   \titlerunning{Temporal  Variation  of the  Hemispheric Solar Rotation}  % title_head in odd pages
   \maketitle

%% The author head (on even pages) and the title head (on odd pages) will be
%% automatically extracted from \author{} and \title{}. Whenever the title is too long,
%% you will be asked to supply a shorter one by inserting either \authorrunning{} or
%% \titlerunning{} before \maketitle. Anyway, you can specify your own heads.
%%
%%
%% Note: In the following text body of your manuscript, please note several differences from
%%       other major journals:
%% (1) \subsection{Please Capitalize the First Letter of Each Notional Word in Subsection Title}
%% (2) Please Capitalize the First Letter of Each Notional Word in all tables' captions

%
%________________________________________________ sections below
%
\section{Introduction}           %% first-level sections will be auto-capitalized
\label{sect:intro}

There are two main  methods  used in investigating the solar
rotation rate: the trace method and the spectroscopic method
\cite{Lucio}. And it is found that the Sun has a higher rotation
rate in the equatorial region: 26 days at the equator while 30~days
at $60^\circ$ latitude \cite{Lawrence2008, Le2007}. More details
about  different measures of the Sun's rotation rate can be found in
the review papers \cite{ Howard1984a, Schroeter1985, Snodgrass1992,
Beck2000, Lucio}. Hoping to reach a more synthetic view of solar
rotation, Heristchi \& Mouradian \cite{Heristchi2009} suggested a
method called global rotation applied to structures of solar
activity. Using this method, they indicated that  individual
structures, local proper motions, meridian drift or differential
rotation could be analyzed together in the considered time.

How the solar differential rotation varies  in a solar cycle as well
as in a long time is still  an  unsolved problem. Li et al.
\cite{li2011a, li2011b} used a continuous complex Morlet wavelet
transformation  to investigate the temporal variations of the
rotational cycle length of daily sunspot areas and daily sunspot
numbers from a global point of view, and indicated that  the
rotational cycle length of the Sun had a secular trend, and the
rotational period had no relation with the Schwabe cycle. Li et al.
\cite{li2011b} pointed out that a lower than average rotation
velocity should statistically appear around the maximum time of
solar activity, while around the minimum time the rotation velocity
was very close to the average. But Gilman \& Howard
\cite{Gilman1984}, Zuccarello \& Zappal\'{a} \cite{Zuccarello} and
Braj\v{s}a et al. \cite{ Brajsa2006} claimed a higher than average
rotation velocity appear in the minimum time of solar activity.

The North-South (N-S) asymmetry in solar activity is an important
part of solar physics. A lot of research has been done based on
various solar activity  indices on the solar surface. More details
about the N-S asymmetry can be found in Vizoso \& Ballester
\cite{Vizoso1990}, Verma \cite{Verma1993}, Carbonell et al.
\cite{Carbonell1993, Carbonell2007}, Li et al. \cite{li2001,
li2002,li2010}, and S\'{y}kora \& Ryb\'{a}k \cite{Sykora2010}.
Besides, the rotational periods are also subjected to a N-S
asymmetry \cite{Temmer2003}. Gilman \& Howard \cite{Gilman1984}
found that in the northern hemisphere the rotation was more
solid-body-like. Javaraiah \& Ulrich \cite{Jav2006} indicated that
there existed difference in the hemispheric rotation rates. Howard
et al. \cite{Howard1984b} analyzed the large spots data and found
that the rotation rate increased less in the northern hemisphere.
Antonucci et al. \cite{Antonucci1990} investigated the rotational
period of the photospheric magnetic field during cycle 21
 and their results showed that the  two hemispheres had different
dominant periods--- 26.9 days for the northern hemisphere and 28.1
days for the southern hemisphere. Also, the result of Temmer et al.
\cite{Temmer2002a, Temmer2002b} concerning the rotational periods of
$H\alpha$ flare and sunspot numbers  accorded  with the periods
found by Antonucci et al. \cite{Antonucci1990}. However, the
observational result of Balthasar et al. \cite{Balthasar1986}
indicated that the sunspots had a little higher rotation rate in the
southern hemisphere by analyzing sunspot groups of all types in the
period 1874-1976. Georgieva \& kirov \cite{Georgieva2003} indicated
that the two hemispheres not only rotated differently but aslo had
different periodicities  in the variations of the rotation
parameters. The N-S asymmetry in hydrogen filament rotation has been
studied by Gigolashvili \cite{Gigolashvili2001} and Gigolashvili et
al. \cite{Gigolashvili2003}. They found that the sign of asymmetry
changed with the hale period, and they suggested that the N-S
asymmetry of the solar rotation might be connected with the N-S
asymmetry of solar activity.

 This work follows the
previous study of  Li et al. \cite{li2011a,li2011b}. We still use
the continuous complex Morlet wavelet transformation to obtain the
rotational signals reflected in the daily hemispheric sunspots'
wavelet power spectrum from a global point of view, and then conduct
further research on temporal variation of the solar  rotation
separately into the northern and southern hemispheres and on their
relationship with the hemispheric solar activity. In addition, we
investigate the N-S asymmetry of the solar rotational cycle length,
including its time-variation, its periodicity,  and also its
relationship with the  N-S asymmetry of solar activity.

% Authors can give a citation as `Michel et al. 1992'.
% You may also use \cite, \citep and \citet for citation, and use Table~1
% or Figure~1 and so forth. Using \ref and \label for cross-references of
% Tables/Figures is a good way in adjusting/adding/removing text, tables or
% figures.

\section{The rotational signal in daily Hemispheric sunspot numbers}
\label{sect:Obs}

\subsection{Data}

The time series data analyzed in our study are:

(1) The daily  northern and southern hemispheric sunspot numbers (
January 1, 1945 to December 31, 2004 )\footnote{
ftp://cdsarc.u-strasbg.fr/pub/cats/J/A+A/447/735/},  compiled by
Temmer et al. \cite{Temmer2006}.

(2) The daily northern and southern hemispheric international
sunspot numbers (January 1, 2005 to December 31, 2010)\footnote{
http://sidc.oma.be/sunspot-data/}. This time series actually
 starts from  January 1, 1992, thus there is an overlapping time span from
January 1, 1992 through  December 31, 2004 with the first time
series. However, in general, the first one renders the second very
well (for details, see Temmer et al. \cite{Temmer2006}).

Figure 1 shows the data and their linear regressions against time
(daily sunspot number on the whole solar disk  at a certain time is
the number in the northern hemisphere plus that in the south at the
same time). It's obvious that the daily sunspot numbers of  the
whole solar disk as well as the northern and southern hemispheres
all have a decrease trend during the time interval considered.

\begin{figure}[!ht]
  \begin{minipage}[t]{0.5\linewidth}
  \centering
   \includegraphics[width=7.0cm, angle=0]{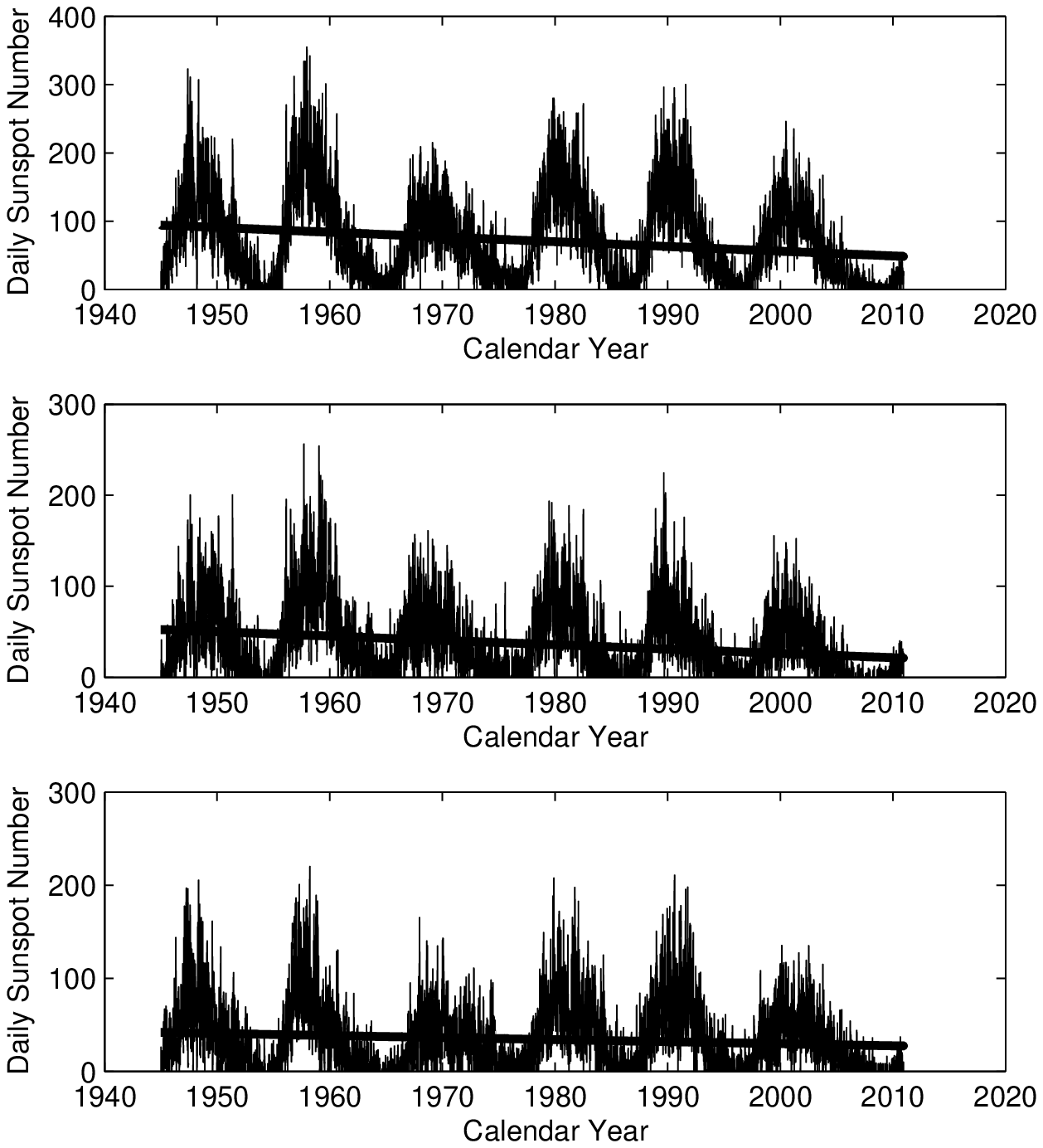}
   \caption{{\small  Daily  sunspot numbers of the whole disk ( top panel),
 the northern hemisphere (middle panel), and the southern
hemisphere (bottom panel) from January 1, 1945 to  December 31,
2010. The thick solid lines are their corresponding linear
regression lines. } }
  \end{minipage}%
  \begin{minipage}[t]{0.55\linewidth}
  \centering
   \includegraphics[width=7.0cm, angle=0]{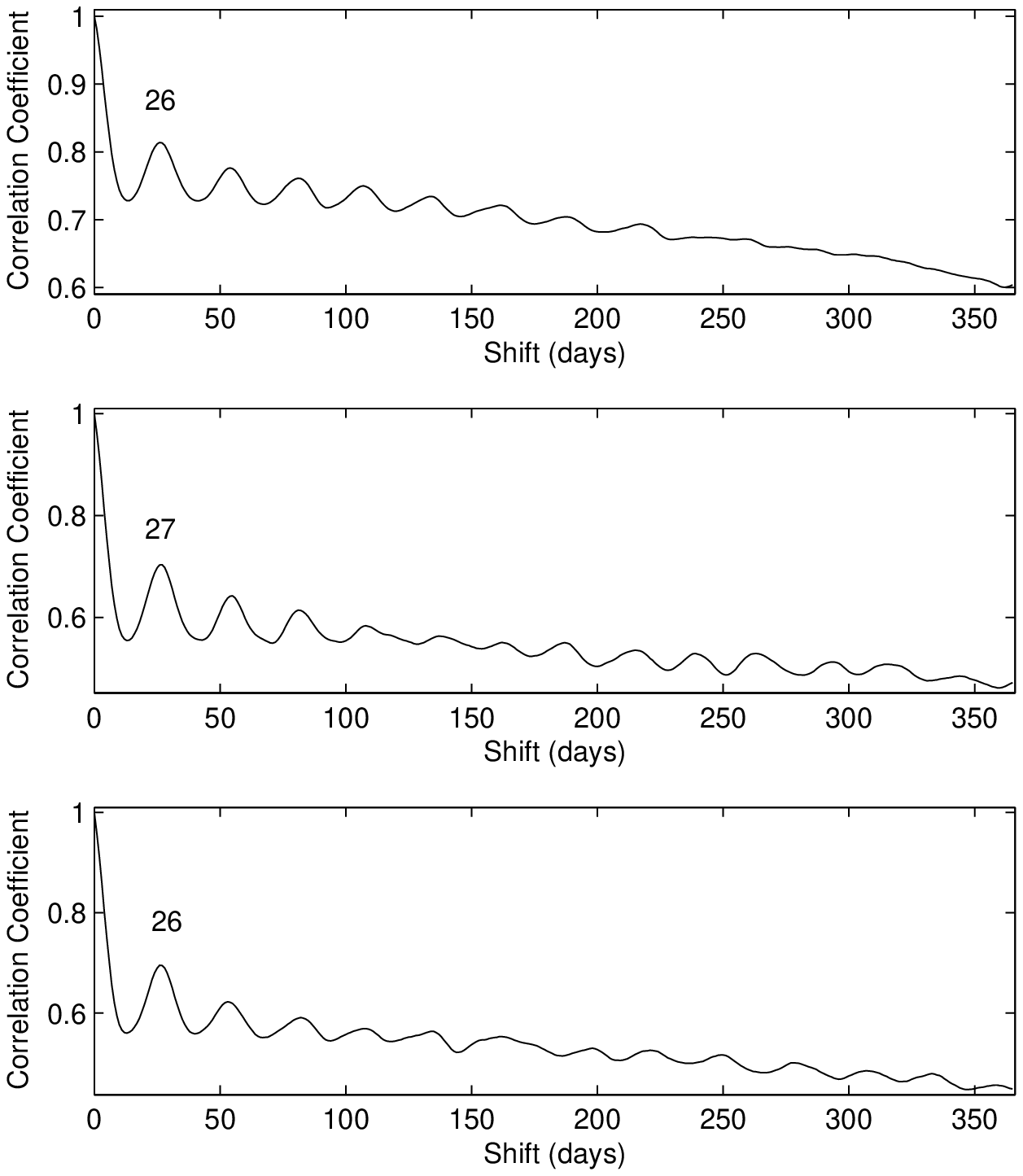}
  \caption{{\small Auto-correlation function of the daily hemispheric sunspot numbers for the
period from  January 1, 1945 to December 31,  2010, plotted up to a
time lag of 365 days. The top panel is for  the whole disk, the
middle panel, for the northern hemisphere, and the bottom panel, for
the southern hemisphere.}}
  \end{minipage}%
  \label{Fig:fig1}
\end{figure}

\begin{figure}[!ht]
\begin{minipage}[t]{0.35\linewidth}
  \centering
   \includegraphics[width=5.3cm, angle=0]{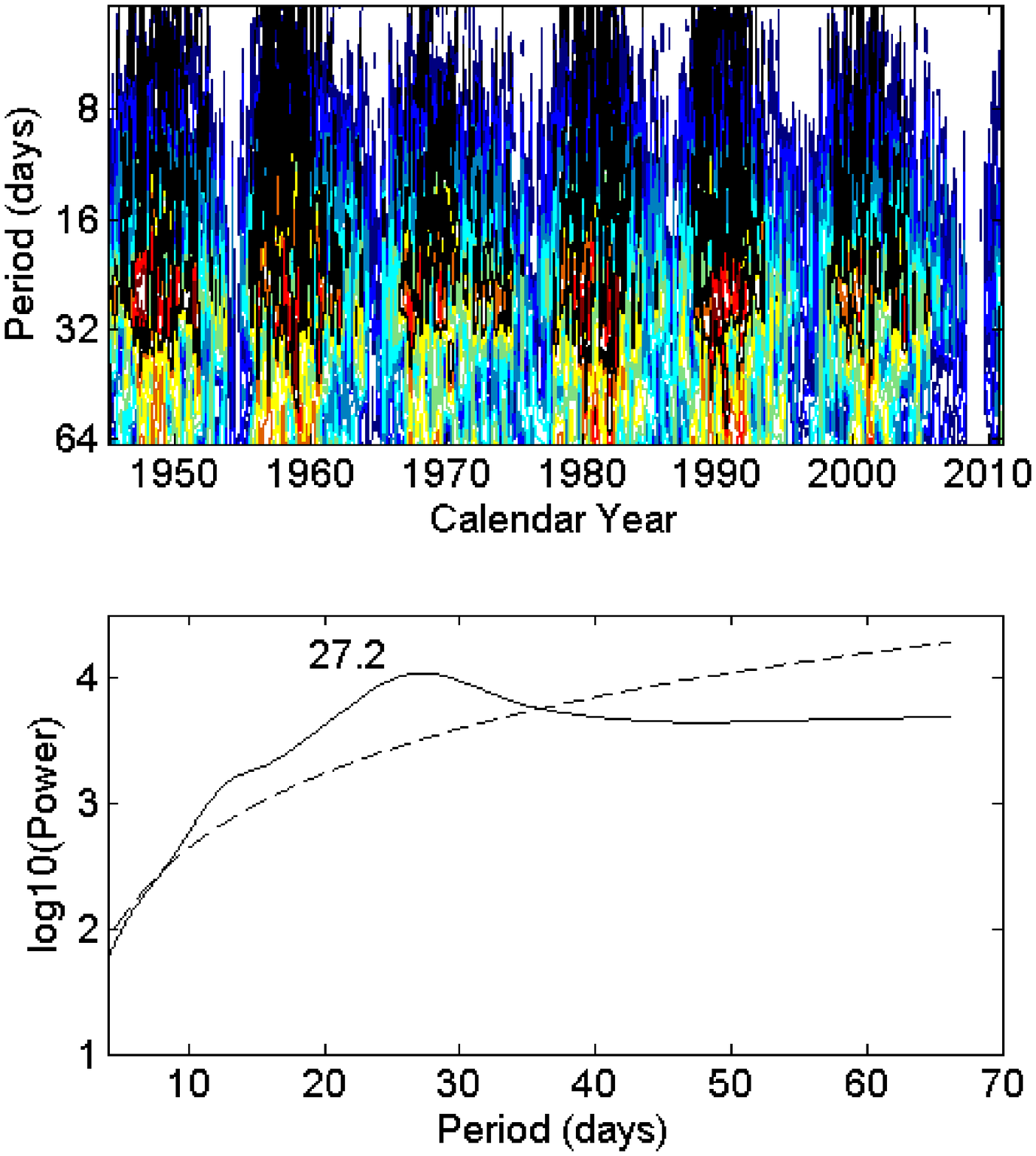}
   \caption{{\small  The top panel: the continuous wavelet power
spectrum of the daily  sunspot numbers. The bottom panel: the global
power spectrum (the solid line) of daily  sunspot numbers. The
dashed line shows the 95\% confidence level.} }
  \end{minipage}%
  \begin{minipage}[t]{0.35\linewidth}
  \centering
   \includegraphics[width=5.3cm, angle=0]{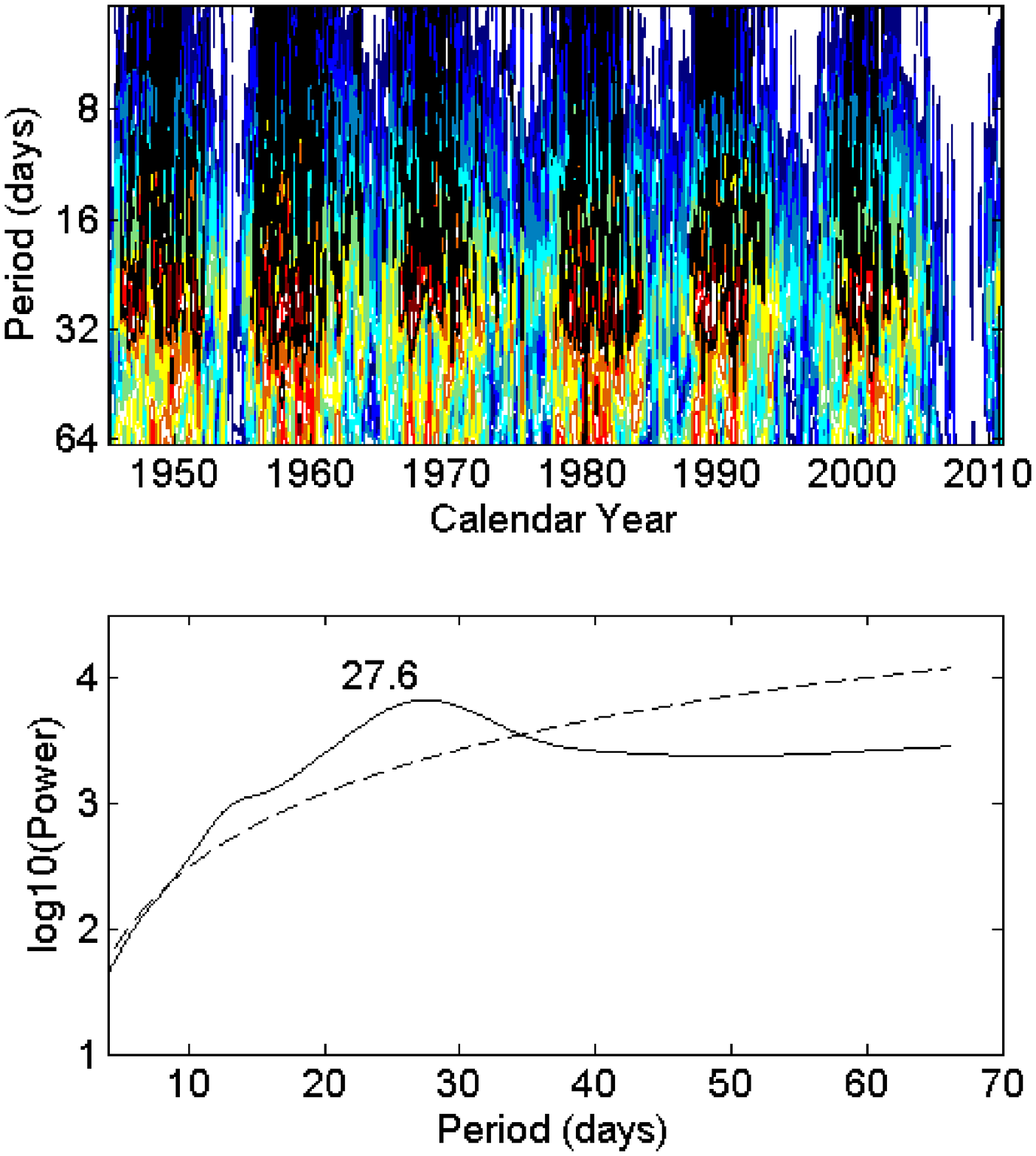}
  \caption{{\small The top panel: the continuous wavelet power spectrum
of the daily northern sunspot numbers. The bottom panel: the global
power spectrum (the solid line) of daily northern sunspot numbers.
The dashed line shows the 95\% confidence level.}}
  \end{minipage}%
 \begin{minipage}[t]{0.35\linewidth}
  \centering
   \includegraphics[width=5.3cm, angle=0]{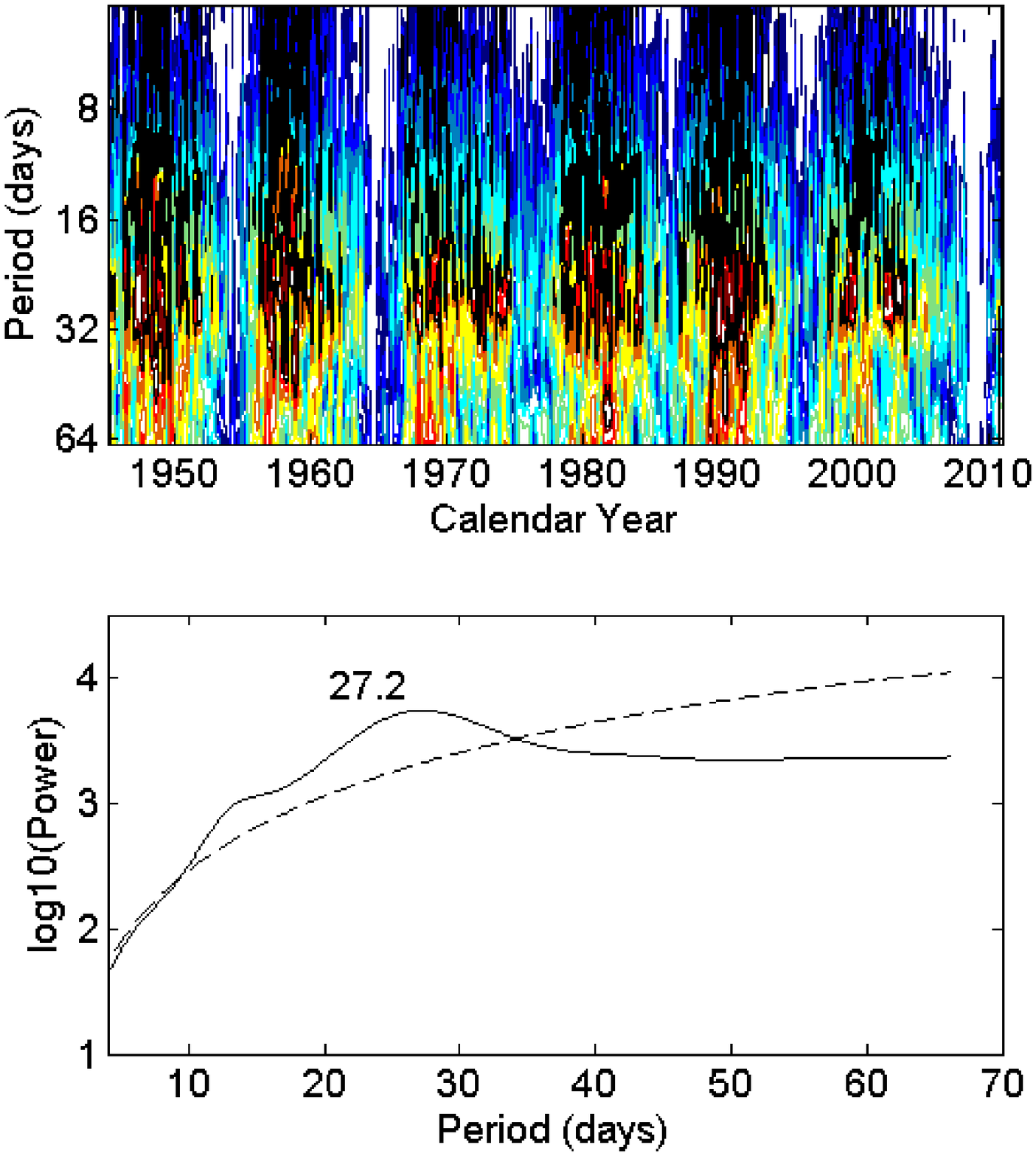}
   \caption{{\small  The top panel: the continuous wavelet power spectrum
of the daily  southern sunspot numbers. The bottom panel: the global
power spectrum (the solid line) of daily southern  sunspot numbers.
The dashed line shows the 95\% confidence level.} }
  \end{minipage}%
  \label{Fig:fig2}
\end{figure}

\subsection{Rotational Period}

The auto-correlation function  is used  here to detect the
periodicity of daily hemispheric sunspot numbers, which is shown in
Figure 2. The auto-correlation coefficients of the daily hemispheric
sunspot numbers  show that the rotational period is the only period
in the time scale shorter than 1 year, whose value is 26, 27, and 26
days for the whole disk, the northern,  and southern hemispheres,
respectively. It means that  the southern hemisphere rotates faster
 over the considered time interval.

We also employ the continuous complex Morlet (dimensionless
frequency $\omega_0$=6) wavelet transformation \cite{Torrence1998}
here to study the periodicity of daily hemispheric sunspot numbers.
 The wavelet anylysis decomposes a transform from a one-dimensional time series into a
two-dimensional time-frequency space. Therefore, this method
determines not only the periodicities of the dominant modes of
variability, but also shows how the modes vary in time
\cite{Torrence1998, Chowdhury2011}. The Morlet wavelet used in the
paper is defined as
\begin{equation}
\Psi_0(\eta)=\pi^{-1/4}e^{i\omega_0\eta}e^{-{\eta^2}/2}
\end{equation}
where $\omega_0$ is the dimensionless frequency and $\eta$ is the
dimensionless time. When using wavelets for feature-extraction
purposes, the Morlet wavelet (with $\omega_0$=6) is a good choice,
since it provides a good balance between time and frequency
localization \cite{Torrence1998, Grinsted2004}.

 As the wavelet is not
completely localized in time, the continuous wavelet transformation
is subject to edge artefacts. It is thus useful to introduce a cone
of influence (COI) in which the transform suffers from these edge
effects (see  the regions outside the thick dashed lines in Figures
8 to 10 as well as in Figure 14). The COI is defined as the wavelet
power for a discontinuity at the edges decreases by a factor
$e^{-2}$. Portions of the transform that are outside the area
encompassed by the time axis and the COI are subject to these edge
effects and are therefore unreliable \cite{Torrence1998,
Grinsted2004, De2004, li2006, Chowdhury2011}.

The significance levels for the wavelet power spectra are calculated
assuming a mean background spectrum modeled with a univariate lag-1
autoregressive process. To determine the significance levels, one
first needs to choose an appropriate background spectrum. For many
 time series, an appropriate background spectrum is either white
noise or red noise. Throughout this paper, the statistical
significance test is carried out by assuming that the noise has a
red spectrum, that is a red noise background is considered. In a red
noise spectrum, the discrete Fourier power spectrum, after
normalizing, is
\begin{equation}
P_k=\frac{1-\alpha^2}{1+\alpha^2-2\alpha\cos(2\pi k/N)}
\end{equation}
where $k=0$, \ldots, $N/2$ is the frequency index, $N$ is the number
of data and $\alpha$ is the  assumed lag-1 autocorrelation. When
$\alpha= 0$, we obtain the white-noise spectrum with an expectation
value of one at all frequencies \cite{Torrence1998,  Chowdhury2011}.

Presented in Figures 3 to 5 are the local wavelet power spectrum and
the global power spectrum of the daily sunspot numbers for the whole
disk, the northern hemisphere, and the southern hemisphere,
respectively. Before performing the wavelet transformation, the raw
data need to be normalized, that is the process of subtracting the
mean value of the data and then dividing by the variance of the
data. As the local wavelet power spectrum shows, the highest power
belt appears around the rotational cycle of the Sun, and it can be
seen clearly around the maximum time of the sunspot cycle. The
figures of the global power spectrum also indicate that the rotation
period is the only period (at the 95\% confidence level), in the
time scale shorter than 64 days. From a global point of view, the
values of the rotational periods
 are 27.2, 27.6, and 27.2 days for the the whole disk, the northern and
southern hemispheres, respectively.

\subsection{Long-term Variations of the Solar Rotation}

\begin{figure}[!ht]
  \begin{minipage}[t]{0.5\linewidth}
  \centering
   \includegraphics[width=7.2cm, angle=0]{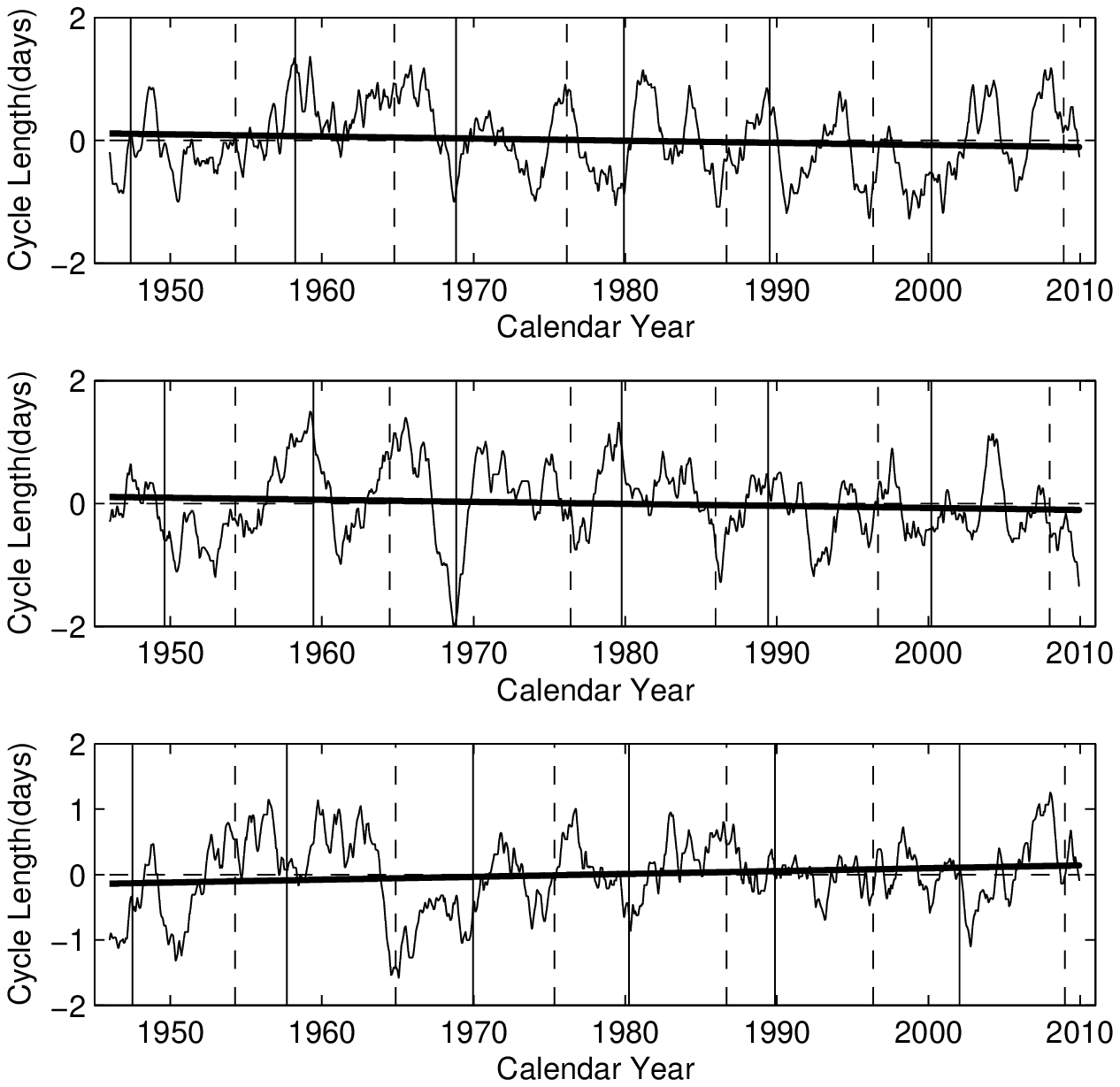}
   \caption{{\small   The period length (the thin lines) of the rotational cycle of daily
 sunspot numbers,
   relative to  mean cycle length. The  thick solid lines show
their secular trends. The vertical solid (dashed) lines indicate the
maximum (minimum) times of sunspot cycles. The three panels are for
the whole disk, the northern hemisphere and the southern one from
top to bottom, respectively.} }
  \end{minipage}%
  \begin{minipage}[t]{0.5\textwidth}
  \centering
   \includegraphics[width=7.2cm, angle=0]{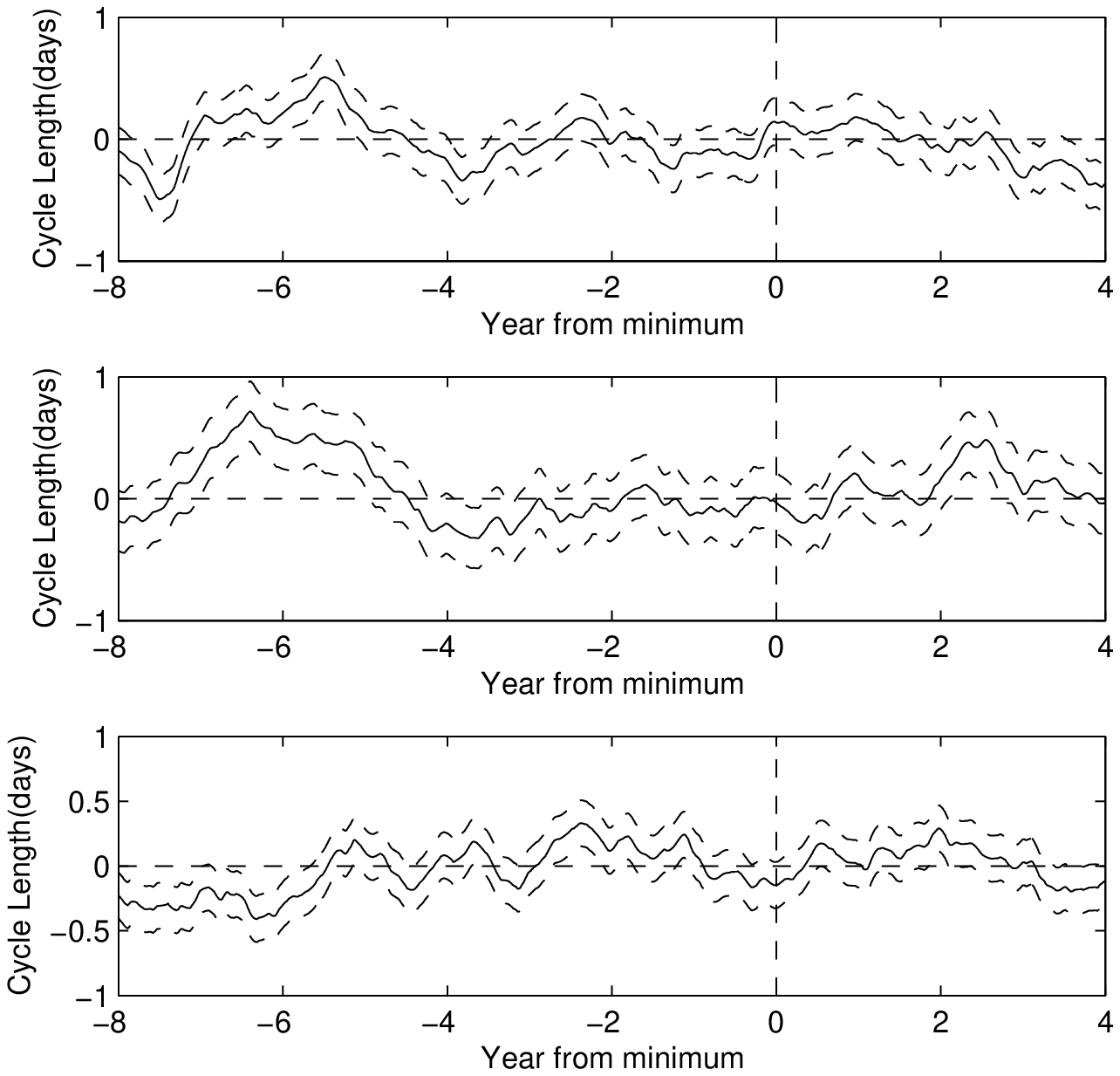}
  \caption{{\small Dependence of the period length (the solid lines) of the rotational cycle
  (relative to mean cycle length) on the phase of the solar
cycle, with respect to the nearest preceding sunspot minimum. The
dashed lines show their corresponding standard errors. The three
panels are for the whole disk, the northern hemisphere and the
southern one from top to bottom, respectively.}}
  \end{minipage}%
  \label{Fig:fig3}
\end{figure}

How the period length of the  rotational cycle (PLRC) changes with
time are presented in figure 6. At a certain time point, the
rotational period (scale) has the highest spectral power among the
period scales of 25 to 31 days (the cycle length of the differential
rotation of sunspots also locates within this range(Temmer et al.
2002b; Yin et al. 2007)) in the local wavelet power spectrum (see
Figs. 3 to 5), upon that the rotational period at each time can be
determined. After that, a 2-year smoothing was introduced to the
obtained temporal variation of the rotational cycle length, and the
new time series and the linear regression lines are given in  Figure
6. The rotational cycle length  of the northern and southern
hemispheres has different varying trends, but as Figure 1 shows, the
hemispheric sunspot numbers have the same decreasing trends. Hence,
we suggest that  the trend of the rotation rate maybe have no direct
relation with the trend of the sunspot
 numbers.

Moreover, as Braj\v{s}a et al. \cite{Brajsa2006} and Li et al.
\cite{li2011b} did, we also investigate the cycle-related variation
of the solar rotation rate, separated into the whole disk, the
northern and southern hemispheres, respectively. As Figure 7 shows,
in the whole disk and the northern hemisphere, a higher than average
velocity appears before the minimum time of solar activity. However,
in the southern hemisphere, the pattern isn't clear seen. Maybe,
it's affected by the
 phase difference between the northern hemisphere and the southern one (see section 3).

\subsection{The Periodicity in the Temporal PLRC }

\begin{figure}[!ht]
  \begin{minipage}[t]{0.35\linewidth}
  \centering
   \includegraphics[width=5.2cm, angle=0]{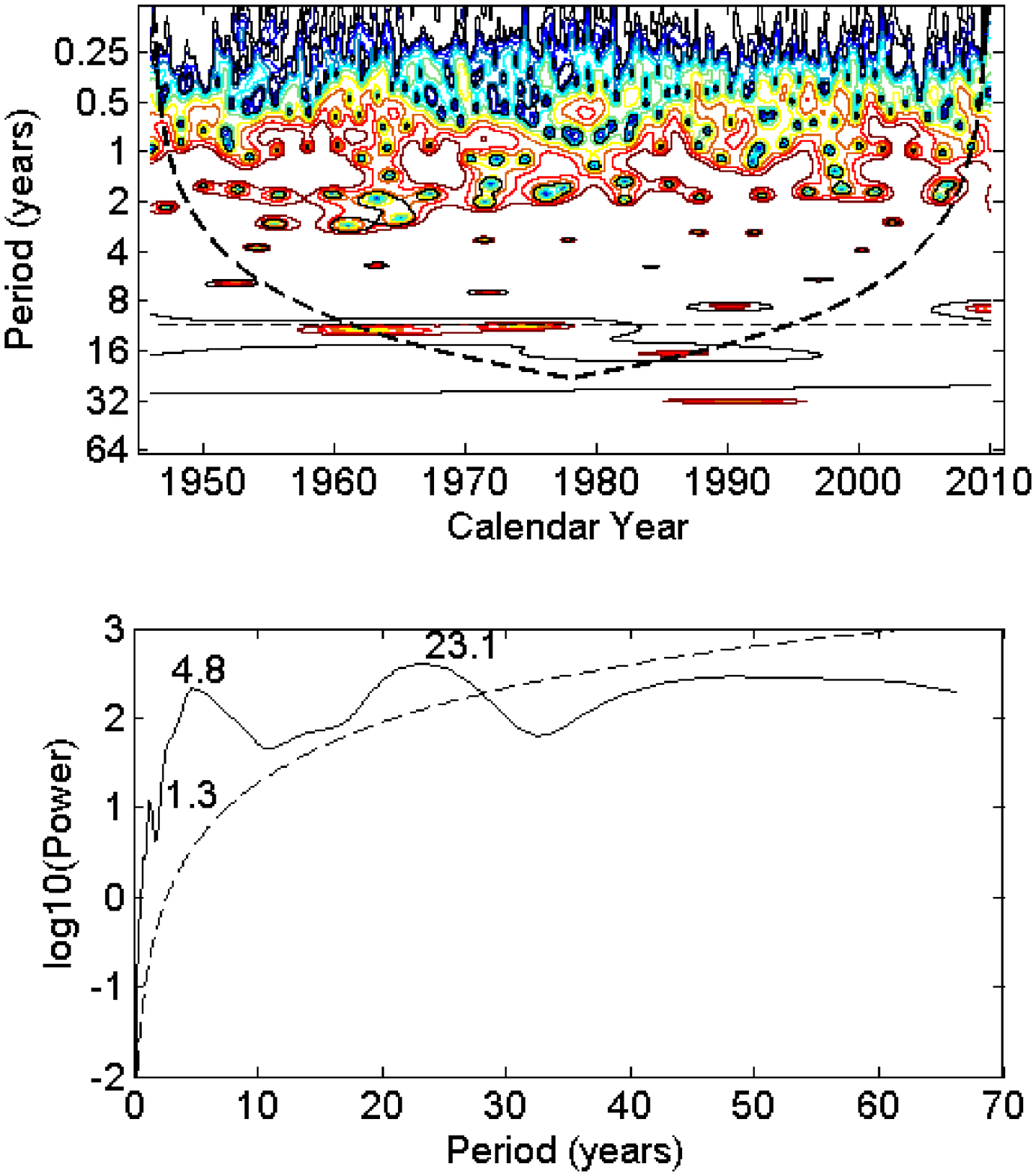}
   \caption{{\small  The top panel: the continuous wavelet power spectrum
of the period length of the rotational cycle of
 the daily  sunspot numbers. The  black solid contours indicate
 the 95\% confidence level. The region below the
thick dashed line indicates the COI where edge effects might distort
the picture \cite{Torrence1998}. The horizontal dashed line stands
for the scale of 11.0 years. The bottom panel: the global power
spectrum (the solid line) of the period length of the rotational
cycle of daily  sunspot numbers. The dashed line shows the 95\%
confidence level.} }
  \end{minipage}%
  \begin{minipage}[t]{0.35\textwidth}
  \centering
   \includegraphics[width=5.2cm, angle=0]{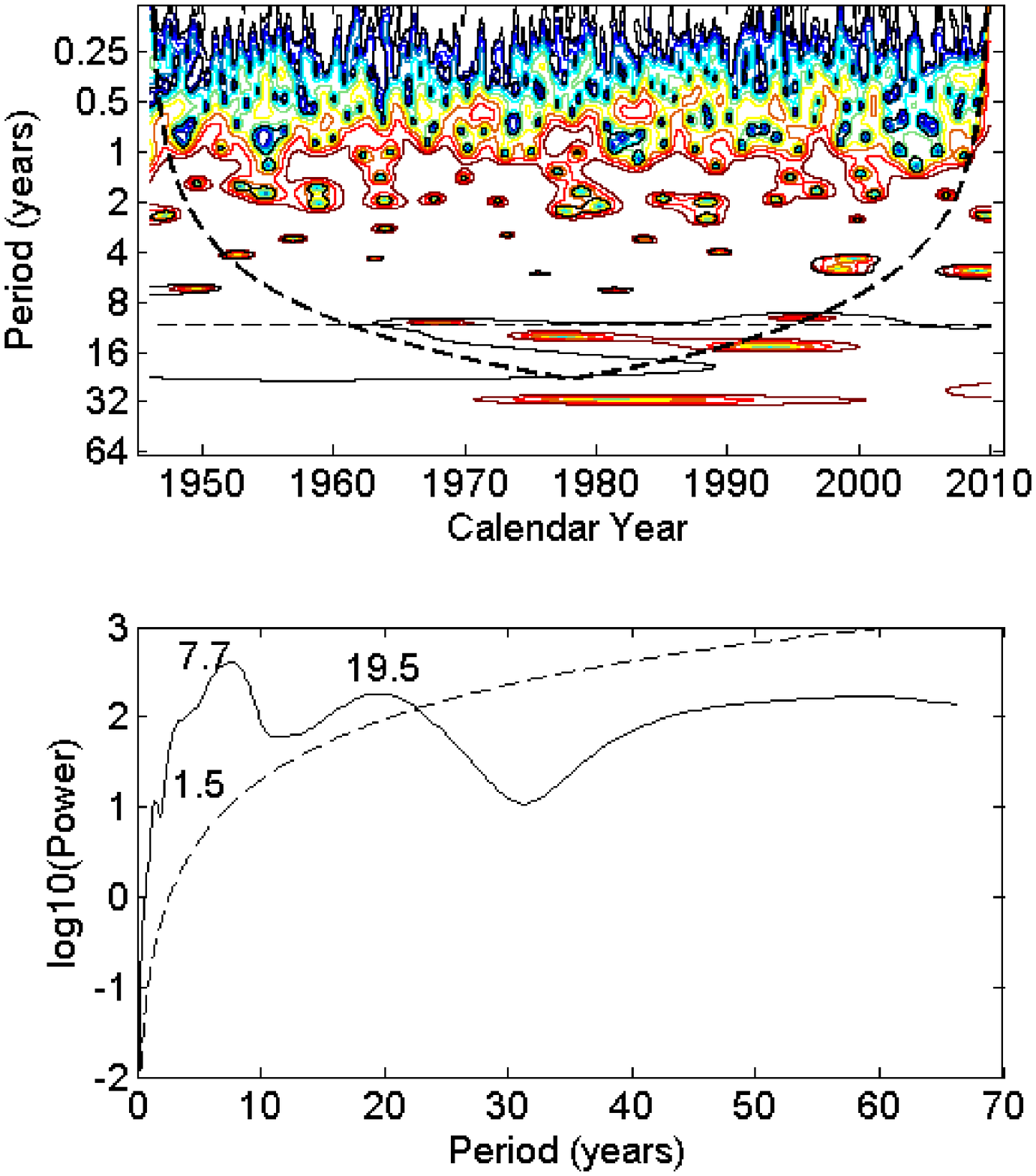}
  \caption{{\small The top panel: the continuous wavelet power spectrum of
the period length of the rotational cycle of
 the daily  northern sunspot numbers. The  black solid contours indicate
 the 95\% confidence level. The region below the
thick dashed line indicates the COI where edge effects might distort
the picture \cite{Torrence1998}. The horizontal dashed line stands
for the scale of 11.0 years. The bottom panel: the global power
spectrum  (the solid line) of  the period length of the rotational
cycle of daily northern  sunspot numbers. The dashed line shows the
95\% confidence level.}}
  \end{minipage}%
 \begin{minipage}[t]{0.35\linewidth}
  \centering
   \includegraphics[width=5.2cm, angle=0]{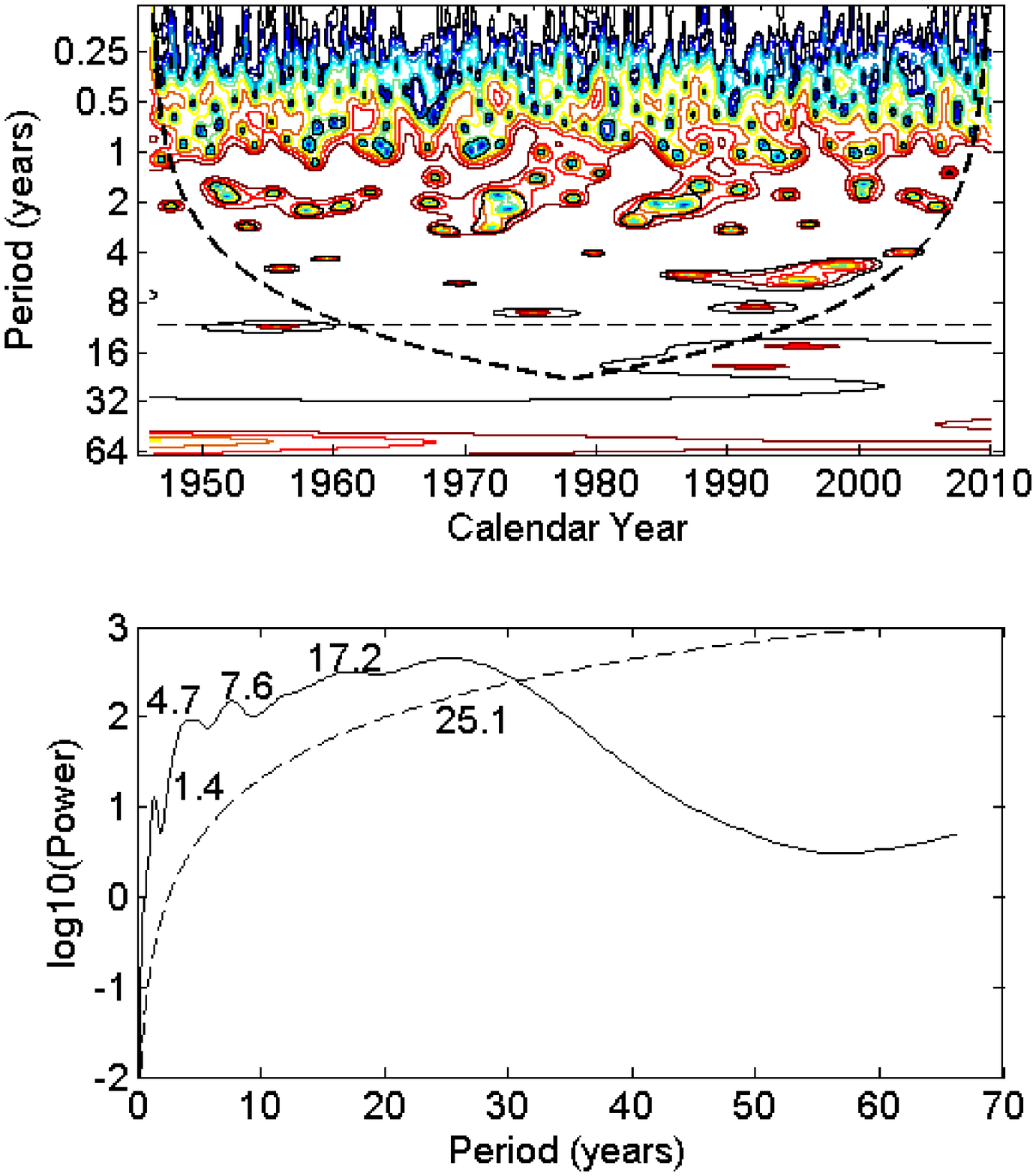}
   \caption{{\small  The top panel: the continuous wavelet power spectrum
   of the period length of the rotational cycle of
 the daily  southern sunspot numbers. The  black solid contours indicate
 the 95\% confidence level. The region below the
thick dashed line indicates the COI where edge effects might distort
the picture \cite{Torrence1998}. The horizontal dashed line stands
for the scale of 11.0 years. The bottom panel: the global power
spectrum  (the solid line) of the period length of the rotational
cycle of daily southern  sunspot numbers. The dashed line shows the
95\% confidence level.} }
  \end{minipage}%
  \label{Fig:fig4}
\end{figure}

For further study, the complex Morlet (dimensionless frequency
$\omega_0$=6) wavelet transformation \cite{Torrence1998} is used
again to investigate  the periodicity in the temporal PLRC of daily
hemispheric sunspot numbers, and the results are represented in
Figures 8 to 10. PLRC is normalized first, too. For PLRC, no
significant period (scale) seems to appear at the 11-year Schwabe
cycle in the whole disk as well as in the northern and southern
hemispheres. This indicates that PLRC might have no relation with
the Schwabe cycle, in agreement with Li et al. \cite{li2011a}.
However, two significant periods of  7.7 and 19.5 years can be seen
for the northern hemisphere, while 4.7, 7.6, 17.2 and 25.1 years for
the southern hemisphere, and 4.8 and 23.1 years for the whole disk
(as the  data are 2-year smoothed, periods less than 2 years are not
reliable).

\section{Relationship of PLRC with solar activity }
\label{sect:relationship}

Figure 11 shows the cross-correlation coefficient between the
smoothed rotational cycle length of the northern and  southern
hemispheres. In the figure, the abscissa indicates the shift of the
northern hemispheric rotational cycle length  with respect to the
southern hemispheric rotational cycle length, with negative values
representing backward shifts. From the figure, one can find that the
northern one lags  the southern one by about 3 years.

\begin{figure}[!ht]
   \centering
   \includegraphics[width=9.0cm, angle=0]{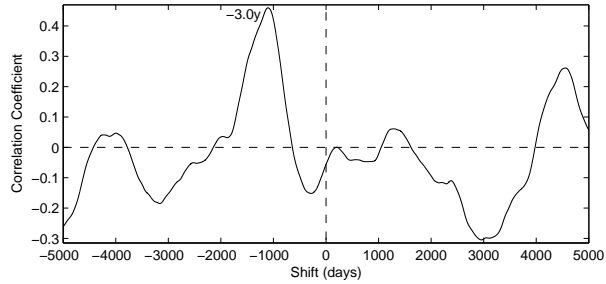}
 \caption{Cross-correlation coefficient between the
smoothed rotational cycle length of the northern and southern
hemispheres, varying with the relative phase shifts between the two.
}
   \label{Fig:fig5}
   \end{figure}

\begin{figure}[!ht]
   \centering
   \includegraphics[width=9cm, angle=0]{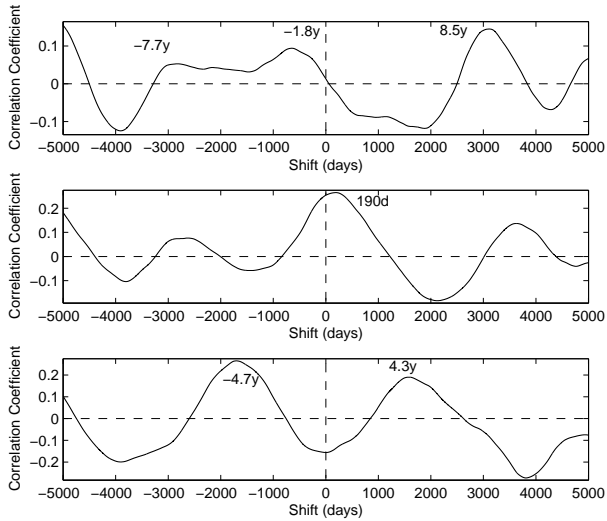}
   \caption{ The cross-correlation coefficient between the
rotational cycle length  and the corresponding 2-year smoothed daily
hemispheric sunspot numbers. The three panels  correspond to the
whole disk, the northern hemisphere and the southern hemisphere from
top to bottom.
    }
   \label{Fig:fig6}
   \end{figure}

Figure 12 shows the cross-correlation coefficient between the
rotational cycle length  and the corresponding 2-year smoothed daily
hemispheric sunspot numbers. In the figure, the abscissa indicates
the shift of the rotational cycle length with respect to the daily
hemispheric sunspot numbers, with negative values representing
backward shifts. From the figure, one can find that rotational cycle
length lags the sunspot numbers by about 1.8 years in the whole
disk, leads by about 190 days in the northern hemisphere, and lags
by about 4.7 years in the southern hemisphere. The three phase
shifts are all different from one another, and the phase shifts in
the whole Sun and in the northern hemisphere are both small,
therefore the solar-cycle related variations of the rotational cycle
length on the whole Sun and in the northern hemisphere look  more
similar with each other.

\section{N-S asymmetry of the rotational cycle length}
\label{sect:asymmetry}

\subsection{N-S Asymmetry }

The N-S asymmetry is calculated traditionally by means of
$Asymmetry= (N - S)/(N + S)$, where $N$ and $S$ stand for the
rotational cycle length (or the 2-year smoothed daily sunspot
numbers) in the northern and southern hemispheres, respectively. The
obtained values (the thin solid lines) and their regression lines
(the thick dashed lines) are plotted in Fig. 13. From the figure,
one can find that the variation trend of N-S asymmetry of the
rotational cycle length displays the same variation trend as the N-S
asymmetry of daily sunspot numbers in the considered time interval.
It means that in the considered time interval  while solar activity
in the southern hemisphere becomes stronger and stronger,  the
southern hemisphere rotates more and more slowly.

\begin{figure}[!ht]
\begin{center}
\includegraphics[width=130mm]{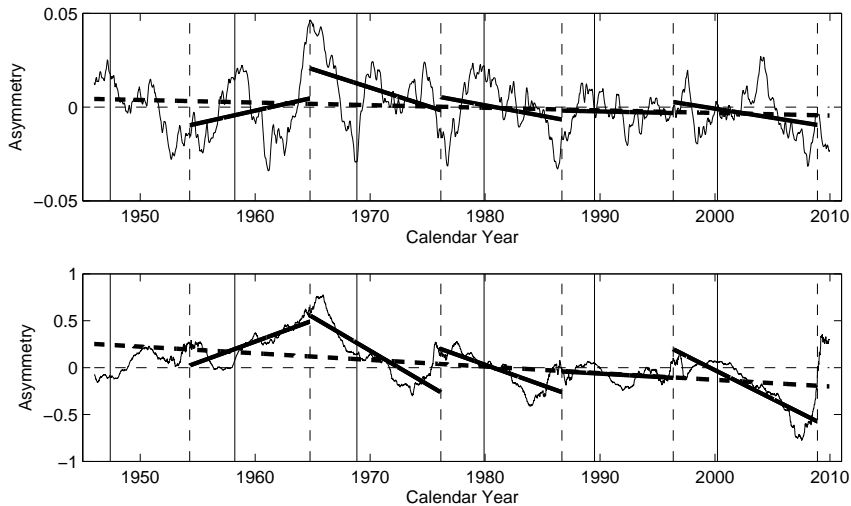}
\caption{\small The  fit of a regression line to the daily value of
the N-S asymmetry of the rotational cycle length (top panel) and
that of the sunspot numbers (bottom panel) for solar cycles 19 to
23. The vertical solid (dashed) lines indicate the maximum (minimum)
times of sunspot cycles.}
   \label{Fig:fig7}
   \end{center}
\end{figure}

As  Vizoso \& Ballester \cite{Vizoso1990},  Ata\c{c} \&
\"{O}zg\"{u}\c{c} \cite{Atac1996}, and Li et al. \cite{li2002} did,
here we fit a straight line to the daily values of the asymmetry,
for each of solar cycles 19 to 23 separately (cycles 18 and 24 are
not a complete cycle in the considered time), starting each cycle
with the time of the minimum between two consecutive cycles (see the
thick solid lines in Fig. 13). The panels show that the slopes of
these fitted straight lines are  positive  for the first cycle, but
negative  for the subsequent four cycles.  Such a positive
(negative) sign for a cycle here in the top panel of Fig. 13 means
that the northern (southern) hemisphere rotates more and more
slowly, related to the southern (northern) hemisphere, when solar
activity is progressing into the cycle. And a positive (negative)
sign for a cycle  in the bottom panel of Fig. 13 means that the
northern (southern) hemispheric solar activity becomes stronger and
stronger, related to the southern (northern) hemispheric solar
activity, when solar activity is progressing into the cycle.
Comparing the two panels of Fig. 13, we can conclude that, in a
solar cycle, when one hemispheric solar activity becomes stronger
and stronger, this hemisphere rotates more and more slowly. Vizoso
\& Ballester \cite{Vizoso1990} proposed a regularity that the slope
of the straight fitted line changes its sign for every four cycles,
and there has been no exception so far. From the preceding
discussion and the regularity, it's  inferred that on cycle 24, the
N-S asymmetry of the hemispheric PLRC should have a positive sign,
accordingly at the beginning of  cycle 24, the northern hemisphere
should have a shorter rotational periods, namely the northern
hemisphere should rotate faster at first, and then it will rotate
more and more slowly.

\begin{figure}[!ht]
   \centering
   \includegraphics[width=11.0cm, angle=0]{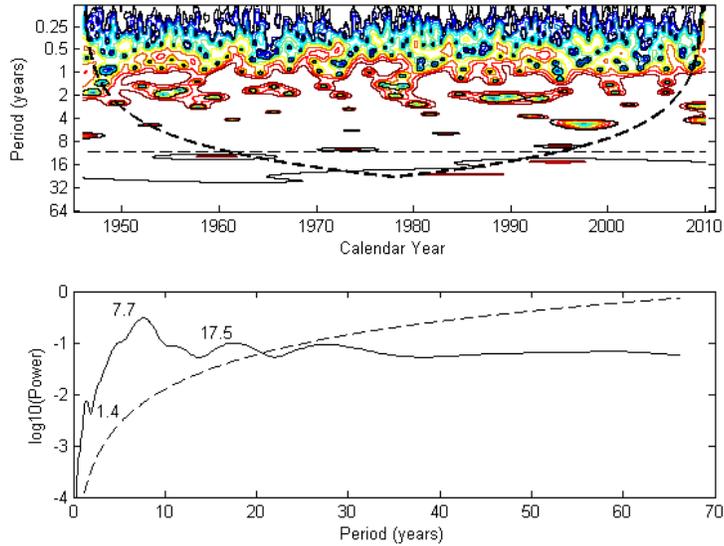}
   \caption{\small The top panel:  the continuous wavelet power spectrum
of the N-S asymmetry of  the rotational cycle length of the daily
hemispheric sunspot numbers. The  black solid contours indicate
 the 95\% confidence level. The region below the
thick dashed line indicates the COI where edge effects might distort
the picture \cite{Torrence1998}. The  horizontal dashed line stands
for the scale of 11.0 years. The bottom panel: the global power
spectrum (the solid line) of the N-S asymmetry of the rotational
cycle length of the daily hemispheric sunspot numbers. The dashed
line shows the 95\% confidence level.
    }
   \label{Fig:fig8}
   \end{figure}

What's more, we investigate the periodicity of the N-S asymmetry of
the daily hemispheric rotational cycle length by using the  complex
Morlet wavelet transformation again (Figure 14), and find that no
significant period (scale) seems to appear at the 11-year Schwabe
cycle, but there are two significant periods at the 95\% confidence
level, whose values are 7.7  and 17.5 years (as the  data are 2-year
smoothed, 1.4-year period is not reliable),  and the 7.7-year period
consists with aforesaid period of the hemispheric rotational cycle
length.
%the latter consists with the period of N-S asymmetry of solar activity index.

\subsection{Relationship of the N-S Asymmetry of PLRC with the N-S Asymmetry of Solar Activity}

\begin{figure}[!ht]
\begin{center}
\includegraphics[width=90mm]{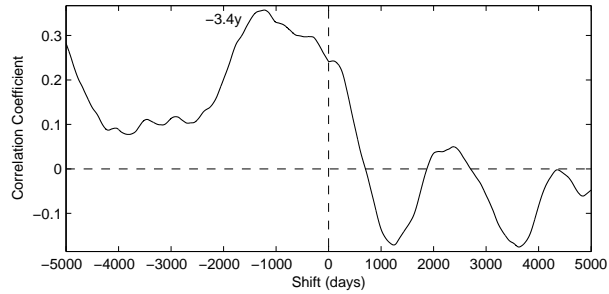}
\caption{ Cross-correlation coefficient between the N-S asymmetry of
the rotational cycle length  and the N-S asymmetry of the daily
hemispheric sunspot numbers, varying with the relative phase shifts
between the two.}
   \label{Fig:fig9}
   \end{center}
\end{figure}

Figure 15 shows the cross-correlation coefficient between the N-S
asymmetry of the rotational cycle length  and the N-S asymmetry of
the daily sunspot numbers. In the figure, the abscissa indicates the
shift of the N-S asymmetry of the rotational cycle length with
respect to the N-S asymmetry of the daily hemispheric sunspot
numbers, with negative values representing backward shifts. As the
figure shows, the two have a high correlation coefficient of 0.24
when there is no shift between the two. And when the  N-S asymmetry
of daily hemispheric sunspot numbers moved backward 3.4 years, the
cross-correlation coefficient reaches its peak.

\section{Conclusions and Discussion }
\label{sect:discussion}

The long-time variations of the solar rotation rate are studied  in
the northern and southern hemispheres respectively through a
continuous wavelet transformation method from a global point of
view, and the main results are listed as follows:

 1. The autocorrelation  function indicates
 that the southern hemisphere rotates faster than the northern hemisphere in
 the considered time interval. Lusting \cite{Lusting}
studied the solar differential rotation by using positions of
sunspots of the years from 1947 to 1981, and found that the southern
hemisphere had a smaller gradient of the differential rotation than
the northern hemisphere had, in  other words,  the southern
hemisphere rotated faster than the northern hemisphere. And from the
Table 1 of Javaraiah et al. \cite{Jav2005}, we may find that the
southern hemisphere indeed had a smaller gradient of differential
rotation. To answer why the southern hemisphere rotates faster, one
need to make further study of the reasons for the difference of the
long-time variations of the gradient of differential rotation in two
hemispheres.

 2. The rotational cycle length of the northern and southern hemispheres has different
 variation trends, while solar activity in the two hemispheres has the same
 variation trend. It means that the long-time variation trend of hemispheric
 rotation rate has no direct relation  with the variation trend of solar activity in the considered time.
  Li et al. \cite{li2011c} pointed out that secular trends of solar
rotation on an average of latitudes or at a certain latitude should
change with latitudes. Thus it may be one possible reason for the
different trends of the two  hemispheres' rotational cycle length
that the sunspots of the northern and southern hemispheres form in
different average latitudes in the considered time. Further research
is needed.

 3. The rotational cycle length of both the hemispheres has no
significant period (scale)  appearing at the 11-year Schwabe cycle,
in accordance with   Li et al. \cite{li2011a}, but both has
significant period of about 7.6 years. And the period of about 7.6
years has been found in the periodicity of the surface equatorial
rotation rate by Javaraiah et al. \cite{Jav2009}.

 4. In the whole disk and the northern hemisphere, a higher than
  average velocity appear before the minimum time of
 solar activity. This may be caused by the phase difference and periodic difference
 between the hemispheric rotational cycle length and the
hemispheric solar activity, and  may be also influenced by
spatial-temporal distribution of the sunspots. The solar-cycle
dependence of the two hemispheres' rotational cycle length is also
different, and this may be the result of  the phase shift between
the northern and southern rotational cycle length, as well as the
phase shift between the northern and southern solar activity.

5. The rotational cycle length of the northern and  southern
hemispheres shows difference in their phases. Additionally, the
phase shifts between the rotational cycle length and the sunspot
numbers in the north, south and the whole disk are different from
one another. Since the relation between the hemispheric rotation and
the hemispheric sunspot activity is complex,  in-depth research is
needed.

6. The N-S asymmetry of the rotational cycle length  has the same
variation trend as the N-S asymmetry of solar activity in a solar
cycle as well as in the considered time interval. The N-S asymmetry
of the rotational cycle length  has two significant periods
--- 7.7 and 17.5 years. Moreover, it has high correlation with
the N-S asymmetry of sunspot activity. On the basis of the
aforementioned characteristic  and the regularity advanced by Vizoso
\& Ballester \cite{Vizoso1990} and Li et al. \cite{li2002}, it's
inferred that the northern hemisphere should rotate faster at the
beginning  of  solar cycle 24.

\normalem
\begin{acknowledgements}
The authors are indebted to Professor Ke-Jun Li for his constructive
ideas and helpful suggestions on manuscript. The authors  wish to
express their gratitude to an anonymous reviewer for his/her careful
reading of the manuscript and valuable comments which improved the
paper considerably. Data used here are all downloaded from Web
sites. The authors express their deep thanks to the staffs of these
Web sites. The wavelet software was provided by C. Torrence and G.
Compo. It is available at URL:
http://paos.colorado.edu/research/wavelets/. This work was funded by
the
 National Natural Science Foundation of China
 (NSFC) under Nos.10873032, 10921303, 11073010 and 40636031, and the
 National Key Research Science Foundation(2011CB811406).
\end{acknowledgements}

%\appendix                  %%appendicial material is supported

%\section{This shows the use of appendix}
%A postscript file is actually an ASCII text file (you may even edit it).
%However, you need to transfer a PDF file or any compressed or packaged
%file in binary mode when using FTP.

\label{lastpage}

\end{document}